\documentclass[conference]{IEEEtran}

\usepackage{cite}
\usepackage{amsmath,amssymb,amsfonts}
\usepackage{graphicx}
\usepackage{textcomp}
\usepackage{xcolor}
\usepackage{epstopdf}
\usepackage{amssymb}
\usepackage{amsmath}
\usepackage{algorithm}
\usepackage{algpseudocode}
\usepackage{amsmath}
\usepackage{multirow}
\usepackage{subfigure}
\usepackage{bm}
\def\BibTeX{{\rm B\kern-.05em{\sc i\kern-.025em b}\kern-.08em
    T\kern-.1667em\lower.7ex\hbox{E}\kern-.125emX}}

\begin{document}

\title{A Probabilistic Bound for Peak Age of Information Guarantee}
\author{
\IEEEauthorblockN{Ailing Zhong$^{1,2,3}$, Zhidu Li$^{1,2,3}$, Tong Tang$^{1,2,3}$, Dapeng Wu$^{1,2,3}$, Ruyan Wang$^{1,2,3}$, Yuming Jiang$^{4}$}
\IEEEauthorblockA{
$^{1}$ School of Communication and Information Engineering, Chongqing University of Posts and Telecommunications, China\\
$^{2}$ Advanced Network and Intelligent Connection Technology Key Laboratory of Chongqing Education Commission of China\\
$^{3}$ Chongqing Key Laboratory of Ubiquitous Sensing and Networking, China\\
$^{4}$ NTNU -- Norwegian University of Science and Technology, Norway\\ 
Email: lizd@cqupt.edu.cn}
}

\maketitle

\begin{abstract}
This paper considers the distribution of a general peak age of information (AoI) model and develops a general analysis approach for probabilistic performance guarantee from the time-domain perspective.
Firstly, a general relationship between the peak AoI and the inter-arrival and service times of packets is revealed.
With the help of martingale theory, a probabilistic bound on the peak AoI is then derived for the general case of endogenous independently and identically distributed increments in information generation and transmission processes.
Thereafter, the application of the obtained bound is illustrated with the M/M/1 and D/M/1 queuing models.
The validity of the proposed bound is finally examined with numerical results.
\end{abstract}


\section{Introduction}

The timeliness or the freshness of information is of great significance for real-time sensing applications, such as factory automation, Metaverse and etc \cite{8469047}.
It is not only related to the end-to-end delay of information, but also relevance with the information generation rate.
Traditional performance metrics such as delay or latency are unable to characterize the information freshness \cite{9014311}.
In this regard, age of information (AoI) and peak AoI are introduced and have become prevailing metrics to quantify the freshness of information \cite{9380899}.
Particularly, the peak AoI is usually applied to characterize the instantaneous information freshness when an update information is received by the destination node \cite{9328793}.

As information may be generated and transmitted through wireless networks, the stochastic properties of wireless channels may have great influence in the peak AoI performance.
Besides, due to the application requirement, the information generation process may also be stochastic.
In this case, it is not possible to provide deterministic peak AoI guarantee to the users.
Consequently, it is necessary to study how to guarantee the peak AoI performance from the probabilistic point of view.

In the literature, most contributions have been devoted to average or expected (peak) AoI analysis and guaranteeing.
In \cite{9322193}, a scheduling policy between associated devices was proposed to reduce the average AoI in Internet of Things.
In \cite{9375486}, service preemption was introduced to improve the average peak AoI for a generate-at-will source node.
In \cite{8813065}, an asymptotic expression of the average peak AoI was derived to determine the superiority of overlay and underlay schemes in a primary IoT system.
In \cite{9681851}, the focus was on the peak AoI performance guarantee for massive machine type communication devices, where a closed-form expression of expected peak AoI was derived by taking the energy harvesting process into account.

To further explore the statistical characteristics of information freshness, some researchers tried to find out the distribution of the peak AoI under some classical queuing models \cite{8820073}.
In \cite{9324753}, a closed-form expression of peak AoI distribution was derived for the M/M/1 and M/D/1 queueing models.
In \cite{9333607}, the distribution of peak AoI was studied based on three different queueing disciplines under the assumption of Bernoulli information packet arrivals.
However, classical queueing model may not be appropriate to the practical network.
Hence, work \cite{9485125} analyzed the peak AoI violation probability with Mellin transform technique in a UAV communication network.
However, as the independent and identically distributed (i.i.d) increments of arrival and service processes were not fully applied, the obtained probabilistic peak AoI bound was loose.
In summary, there is still a gap when exploiting the statistical characteristics of peak AoI in the analysis, particularly for stochastic peak AoI guarantee.

Motivated by this, we apply a time-domain analytical approach for probabilistic peak AoI guarantee analysis under a general setting.
Specifically, without making any assumptions on the information generation and transmission processes, a general peak AoI model is formulated with only inter-arrival time and service time of a packet.
By resorting to the endogenous i.i.d increments in information generation and transmission processes, a probabilistic peak AoI bound is derived based on martingale theory.
Thereafter, the application of the proposed analytical approach and obtained bound is exemplified with two specific cases, where upper bounds on peak AoI violation probability and on average peak AoI are derived.
Finally, the validity and implication of the obtained bounds are examined with numerical results and comparison with the exact results obtained from queueing theory analysis.

The remaining of this paper is organized as follows.
Section II derives the general probabilistic bound of the peak AoI.
Case analysis under specific scenarios is presented in Section III.
In Section IV, numerical results are provided and discussed.
Section V concludes the paper.

\section{Probabilistic Bound of Peak AoI}
In this section, a general probabilistic bound of peak AoI is derived.
Throughout this paper, we use $f_X (\cdot)$ and $F_X (\cdot)$ to represent the probability density function (PDF) and the cumulative distribution function (CDF) of random variable $X$, respectively.
Also, let ${\rm E}[\cdot]$ denote the expectation function.

\subsection{Definition and Decoupling of Peak AoI}
\begin{figure}[t!]
\centerline{\includegraphics[height=2.2in,width=3.2in]{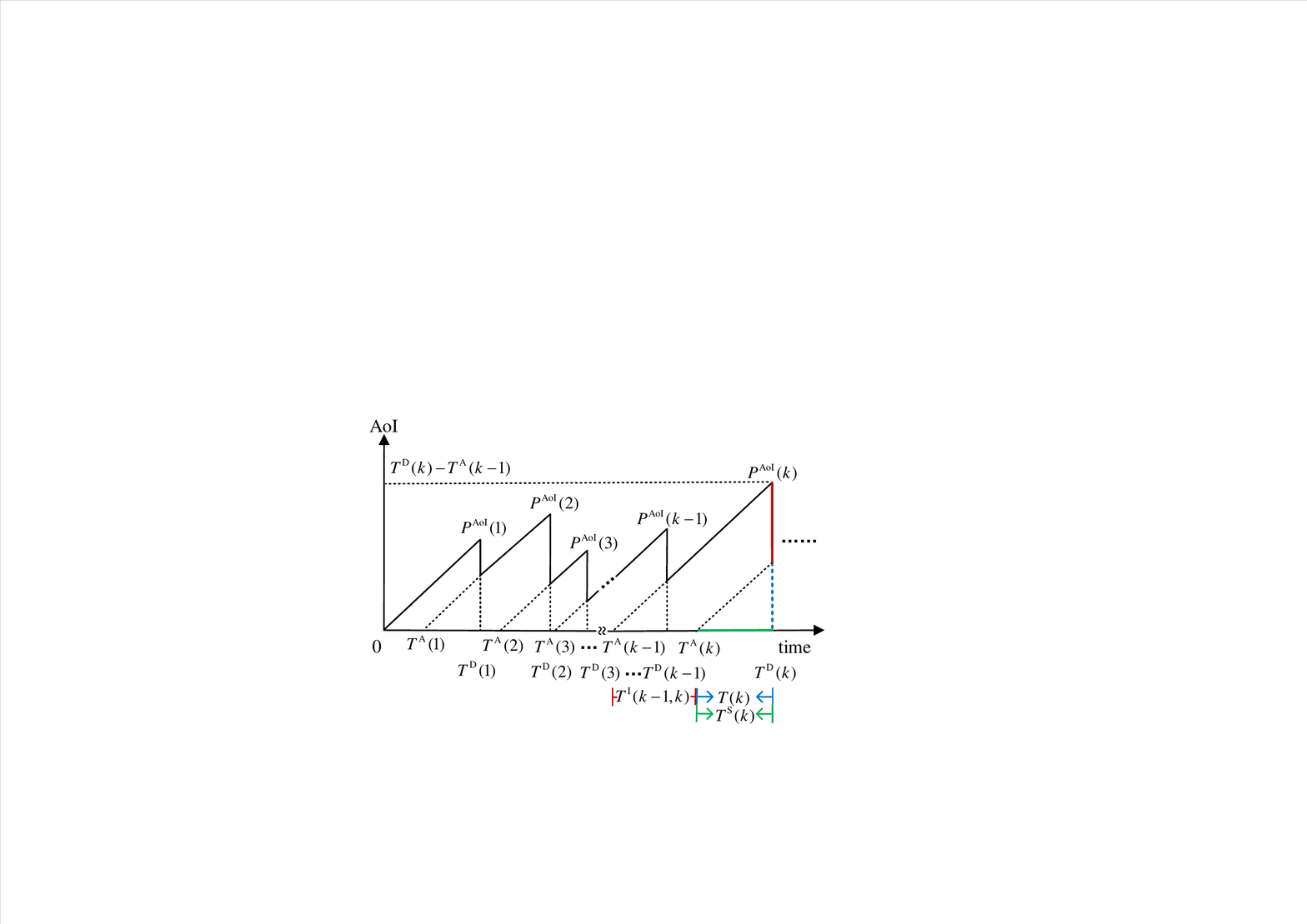}}
\caption{Evolution of the AoI for status-update data packets.}
\label{fig1}
\end{figure}

As depicted in Fig. 1, the peak AoI is defined as the AoI immediately before an information update. Specifically, the peak AoI corresponding to the $k$th packet, denoted as ${P^{{\rm{AoI}}}}(k)$, can be written as:
\begin{equation}
{P^{{\rm{AoI}}}}(k) = {T^{\rm{D}}}(k) - {T^{\rm{A}}}(k - 1)
\end{equation}
where ${T^{\rm{A}}}(k)$ and ${T^{\rm{D}}}(k)$ denote the arrival time and the departure time of the $k$th packet, respectively.
Without loss of generality, we set ${T^{\rm{A}}}(0)=0$ \cite{9485125}.

Under the first-come-first-served (FCFS) policy, the departure time of the $k$th packet can be expressed as \cite{time}:
\begin{equation}
{T^{\rm{D}}}(k) = \mathop {\max }\limits_{{\rm{1}} \le j \le k} \{ {{T^{\rm{A}}}(j) + {T^{\rm{SC}}}(j,k)} \}
\end{equation}
where ${T^{\rm{SC}}}(j,k)$ denotes the cumulative service time from the $j$th packet to the $k$th packet ($j<k$), and there holds
\begin{equation}
{T^{\rm{SC}}}(j,k){\rm{ = }}\sum\limits_{n = j}^k {{T^{\rm{S}}}(n)}
\end{equation}
Here, ${T^{\rm{S}}}(n)$ denotes the service time of the $n$th packet.
It is worth mentioning that the total sojourn time $T(k)$ for any packet is not less than $T^{\rm{S}}$.

Similarly, let ${T^{\rm{I}}}(k - 1,k) = {T^{\rm{A}}}(k) - {T^{\rm{A}}}(k - 1)$ denote the inter-arrival time between the $(k-1)$th packet and the $k$th packet for $k = 1,2, \cdots $.
Then, the inter-arrival time between the $j$th packet and the $k$th packet can be written as:
\begin{equation}
{T^{\rm{I}}}(j,k) = \sum\limits_{n = j{\rm{ + 1}}}^k {{T^{\rm{I}}}(n - 1,n)}
\end{equation}

We assume that the inter-arrival time ${T^{\rm{I}}}(k - 1,k)$ and the service time ${T^{\rm{S}}}(k)$ are both i.i.d for each packet, respectively. In addition we assume the following stability condition:
\begin{equation}
  \begin{aligned}
\rm{E}[T^{\rm{I}}(k - 1,k)] \ge \rm{E}[{T^{\rm{S}}}(k)]
  \end{aligned}
\end{equation}

By integrating  Eqs. (2), (3) and (4), the peak AoI in Eq. (1) can be further expressed as
\begin{equation}
  \begin{aligned}
&{P^{{\rm{AoI}}}}(k) = \mathop {\max }\limits_{1 \le j \le k} \{ {T^{\rm{A}}}(j) + {T^{\rm{S}}}(j,k) - {T^{\rm{A}}}(k - 1)\} \\
& = \max {\rm{\{ }}\mathop {\max }\limits_{1 \le j \le k - 1} {\rm{\{ }}{T^{\rm{A}}}(j) - {T^{\rm{A}}}(k - 1) + {T^{\rm{SC}}}(j,k){\rm{\} }},\\
&~~~~~~~~~~~~~~~~~~~~~~~~~~{T^{\rm{A}}}(k) - {T^{\rm{A}}}(k - 1) + {T^{\rm{S}}}(k){\rm{\} }}\\
& = \max {\rm{\{ }}\mathop {\max }\limits_{1 \le j \le k - 1} {\rm{\{ }}{T^{\rm{SC}}}(j,k) - {T^{\rm{I}}}(j,k - 1){\rm{\} }},\\
&~~~~~~~~~~~~~~~~~~~~~~~~~~{T^{\rm{I}}}(k - 1,k) + {T^{\rm{S}}}(k){\rm{\} }}\\
& = \max {\rm{\{ }}\mathop {\max }\limits_{1 \le j \le k - 1} {\rm{\{ }}{T^{\rm{SC}}}(j,k - 1) - {T^{\rm{I}}}(j,k - 1){\rm{\} }} + {T^{\rm{S}}}(k),\\
&~~~~~~~~~~~~~~~~~~~~~~~~~~~~~~~~~~~~~~~~~~{T^{\rm{I}}}(k - 1,k) + {T^{\rm{S}}}(k){\rm{\} }}\\
& = \max \{ \max \{ \mathop {\max }\limits_{1 \le j \le k - 2} (\sum\limits_{n = j}^{k - 2} {{T^{\rm{S}}}(n) - \sum\limits_{n = j + 1}^{k - 1} {{T^{\rm{I}}}(n - 1,n)} } ),0\} \\
&~~~~~~~~~~~~~~~~~~~~~~~~~~~~ + {T^{\rm{S}}}(k - 1),{T^{\rm{I}}}(k - 1,k)\}  + {T^{\rm{S}}}(k)\\
& = \max \{ \max \{ \mathop {\max }\limits_{1 \le j \le k - 2} (\sum\limits_{n = j}^{k - 2} {[{T^{\rm{S}}}(n) - {T^{\rm{I}}}(n,n + 1)]} ),0\} \\
&~~~~~~~~~~~~~~~~~~~~~~ + {T^{\rm{S}}}(k - 1),{T^{\rm{I}}}(k - 1,k)\}  + {T^{\rm{S}}}(k)
  \end{aligned}
\end{equation}
From Eq. (6), it is worth noting that the peak AoI is affected by not only the inter-arrival time ${T^{\rm{I}}}(k - 1,k)$ but also the service time ${T^{\rm{S}}}(k)$.
Hence, it is impossible to provide deterministic peak AoI guarantee if ${T^{\rm{I}}}(k - 1,k)$ or ${T^{\rm{S}}}(k)$ is random.
In what follows, the peak AoI performance is analyzed from a probabilistic bound or violation probability point of view.

\subsection{Peak AoI Violation Probability}
Let $ Z_k= {T^{\rm{S}}}(k)$, $Y_k= {T^{\rm{I}}}(k - 1,k)$ and introduce an auxiliary parameter $B = \max \{ \mathop {\max }\limits_{1 \le j \le k - 2} (\sum\limits_{n = j}^{k - 2} {[ {{Z_n} - {Y_{n + 1}}} ]} ),0\} $, Eq. (6) can be further simplified as
\begin{equation}
  \begin{aligned}
{P^{{\rm{AoI}}}}(k) = \max \{ {B + {Z_{k - {\rm{1}}}},{Y_k}} \} + {Z_k}
  \end{aligned}
\end{equation}

Since ${T^{\rm{I}}}(k - 1,k)$ and ${T^{\rm{S}}}(k)$ are i.i.d parameters, $Z_k$ and $Y_k$ are both i.i.d for $\forall k=1,2,\cdot\cdot\cdot$.
For a given peak AoI threshold $d$, the corresponding violation probability can be expressed as
\begin{equation}
  \begin{aligned}
&\Pr \{ {P^{{\rm{AoI}}}}(k) > d\} \\
&= \Pr \{ \max \{ B + {Z_{k - 1}},{Y_k}\}  + {Z_k} > d\} \\
&= 1 - \Pr \{ \max \{ B + {Z_{k - 1}},{Y_k}\}  + {Z_k} \le d\} \\
&= 1 - \int_0^d {\Pr \{ \max \{ B + {Z_{k - 1}},{Y_k}\}  \le x\} {f_Z}(d - x)dx}
  \end{aligned}
\end{equation}

In order to further reveal the probabilistic characteristics of the peak AoI, a lemma is introduced in the following.

\emph{\textbf{Lemma 1}}: For $\forall k=1,2,\cdot\cdot\cdot$, there always holds
\begin{equation}
  \begin{aligned}
\Pr \{ \max \{ B + {Z_{k - 1}},{Y_k}\}  \le x\} \ge \Pr \{ B + {Y_k} \le x\}
  \end{aligned}
\end{equation}

\emph{Proof}: We prove this lemma through taking the following two cases into account.

\emph{Case 1}: If $B + {Z_{k - 1}} \ge {Y_k}$, we have
\begin{equation}
  \begin{aligned}
\Pr \{ \max \{ B + {Z_{k - 1}},{Y_k}\}  \le x\} &= \Pr \{ B + {Z_{k - 1}} \le x\} \\
 &= \int_0^x {{F_{Z}}(y)} {f_B}(x - y)dy\\
& \mathop  \ge \limits^{(a)} \int_0^x {{F_Y}(y)} {f_B}(x - y)dy\\
 &= \Pr \{ B + {Y_k} \le x\}
  \end{aligned}
\end{equation}
Here, step (a) holds because $\int_0^\infty  {( {1 - {F_Y}(y)} )dy}  = {\rm E}[Y] \ge {\rm E}[Z] = \int_0^\infty  {( {1 - {F_Z}(y)} )dy} $, and both $F_Z(y)$ and $F_Y(y)$ are increasing functions, where, with the stability condition, ${\rm E}[Y] \ge {\rm E}[Z]$, meaning the average inter-arrival time is not less than the average service time.

\emph{Case 2}: If $B + {Z_{k - 1}} \le {Y_k}$, since $B \ge 0$ by definition, we have
\begin{equation}
  \begin{aligned}
\Pr \{ \max \{ B + {Z_{k - 1}},{Y_k}\}  \le x\} &= \Pr \{ {Y_k} \le x\} \\
 &\ge \Pr \{ B + {Y_k} \le x\}
  \end{aligned}
\end{equation}
Thus, $\Pr \{ \max \{ B + {Z_{k - 1}},{Y_k}\}  \le x\} \ge \Pr \{ B + {Y_k} \le x\}$ always holds, which completes the proof. $\hfill\blacksquare$

Applying Lemma 1 in Eq. (8), the peak AoI violation probability can be further derived as

\begin{equation}
  \begin{aligned}
&\Pr \{ {P^{{\rm{AoI}}}}(k) > d\} \\
&= 1 - \int_0^d {\Pr \{ \max \{ B + {Z_{k - 1}},{Y_k}\}  \le x\} {f_Z}(d - x)dx} \\
&\le 1 - \int_0^d {\Pr \{ B + {Y_k} \le x\} {f_Z}(d - x)dx} \\
& \mathop  = \limits^a 1 - \int_0^d {\int_0^x {\Pr \{ B \le y\} {f_Y}(x - y)dy} {f_Z}(d - x)dx}
  \end{aligned}
\end{equation}
Here, step (a) holds since $B$ is independent with $Y_k$.
For a given data update policy, the characteristic information of inter-arrival time and that of service time of the packets are available.
Then, an upper bound on the peak AoI violation probability is readily found if the probabilistic characteristics of the auxiliary parameter $B$ can be derived.
Specifically, a probabilistic upper bound of $B$ is derived and summarized as the following lemma.

\emph{\textbf{Lemma 2}}: For two independent sets of random variables $\{Y_n\}$ and $\{Z_n\}$, where $n=1,2,\cdot\cdot\cdot$.
In each set, all the random variables are i.i.d.
Let $B = \max \{ \mathop {\max }\limits_{1 \le j \le k - 2} (\sum\limits_{n = j}^{k - 2} {[ {{Z_n} - {Y_{n + 1}}} ]} ),0\} $. If ${\rm{E}}[{e^{\theta {Z_1}}}]{\rm{E}}{[^{ - \theta {Y_1}}}] \le {\rm{1}}$, there  holds
\begin{equation}
  \begin{aligned}
&\Pr \{ {B \le y} \} \ge  {1-e^{ - \theta y}}
  \end{aligned}
\end{equation}
for all $\theta\geq 0$ and $y>0$.

\emph{Proof}:
For any $y>0$, we have
 \begin{equation}
  \begin{aligned}
&\Pr \{ B > y\} \\
&=1-\Pr \{ B \le y\} \\
&= 1-\Pr \{ \max \{ {\mathop {\max }\limits_{1 \le j \le k - 2} ( {\sum\limits_{n = j}^{k - 2} {[ {{Z_n} - {Y_{n + 1}}} ]} } ),0} \} \le y\} \\
&\le 1-\Pr \{ \mathop {\max }\limits_{1 \le j \le k - 2} ( {\sum\limits_{n = j}^{k - 2} {[ {{Z_n} - {Y_{n + 1}}} ]} } ) \le y\} \Pr \{ 0 \le y\} \\
&=1- \Pr \{ \mathop {\max }\limits_{1 \le j \le k - 2} ( {\sum\limits_{n = j}^{k - 2} {[ {{Z_n} - {Y_{n + 1}}} ]} } ) \le y\}\\
&=\Pr \{ \mathop {\max }\limits_{1 \le j \le k - 2} ( {\sum\limits_{n = j}^{k - 2} {[ {{Z_n} - {Y_{n + 1}}} ]} } ) > y\}
  \end{aligned}
\end{equation}

Then, let ${V_j} = {e^{\theta ( {{Z_{(k - j - 2,k - 2)}} - {Y_{(k - j - 2,k - 2)}}} )}}$, i.e., ${V_j} = {e^{\theta ( {{Z_{k - j - 2}} - {Y_{k - j - 2}} + {Z_{k - j - 1}} - {Y_{k - j - 1}} +  \cdots  + {Z_{k - 2}} - {Y_{k - 2}}} )}}$.
There holds:
\begin{equation}
  \begin{aligned}
&{V_{j + 1}} = {e^{\theta ( {{Z_{(k - j - 3,k - 2)}} - {Y_{(k - j - 3,k - 2)}}} )}}\\
&~~~~~~= {e^{\theta ( {{Z_{k - j - 3}} - {Y_{k - j - 3}} + {Z_{k - j - 2}} - {Y_{k - j - 1}} +  \cdots  + {Z_{k - 2}} - {Y_{k - 2}}} )}}\\
&~~~~~~= {V_j}{e^{\theta ( {{Z_{k - j - 3}} - {Y_{k - j - 3}}} )}}
  \end{aligned}
\end{equation}
Since $Y_k$ and $Z_k$ both have i.i.d increments, then we have:
\begin{equation}
  \begin{aligned}
&{\rm{E}}[{V_{j + 1}}|{V_1},{V_2}, \cdots ,{V_j}]\\
&= {\rm{E}}[{V_j}{e^{\theta ( {{Z_{k - j - 3}} - {Y_{k - j - 3}}} )}}|{Z_{k - 2}},{Z_{k - 3}}, \cdots ,{Z_{k - j - 2}},\\
&~~~~~~~~~~~~~~~~~~~~~~~~~~~~~~~{Y_{k - 2}},{Y_{k - 3}}, \cdots ,{Y_{k - j - 2}}]\\
&\mathop  = \limits^a {\rm{E}}[{V_j}|{Z_{k - 2}},{Z_{k - 3}}, \cdots ,{Z_{k - j - 2}},{Y_{k - 2}},{Y_{k - 3}}, \cdots ,{Y_{k - j - 2}}]\\
&~~~{\rm{   E}}[{e^{\theta {Z_{k - j - 3}}}}]{\rm{E}}{[^{ - \theta {Y_{k - j - 3}}}}]\\
&\mathop  = \limits^b {V_j}{\rm{E}}[{e^{\theta {Z_1}}}]{\rm{E}}{[^{ - \theta {Y_1}}}]\\
&\mathop  \le \limits^c {V_j}
  \end{aligned}
\end{equation}
Here, step (a) holds because ${Z_{k - i - 3}}$ and ${Y_{k - i - 3}}$ are independent each other, and also independent of $\{ {{Z_{k - 2}},{Z_{k - 3}}, \cdots ,{Z_{k - j - 2}},{Y_{k - 2}},{Y_{k - 3}}, \cdots ,{Y_{k - j - 2}}} \}$.
Step (b) holds because process $Y_k$ and $Z_k$ both have identical increments, i.e., for the random service time, we have:
\begin{equation}
{\rm{E}}[{e^{\theta {Z_{k - j - 3}}}}] = {\rm{E}}[{e^{\theta {Z_{1}}}}] = {\rm{E}}[{e^{\theta {T^{\rm{S}}}(1)}}]
\end{equation}
Correspondingly, for the random inter-arrival time, there holds:
\begin{equation}
{\rm{E}}[{e^{ - \theta {Y_{k - j - 3}}}}] = {\rm{E}}[{e^{ - \theta {Y_{1}}}}] = {\rm{E}}[{e^{ - \theta {T^{\rm{I}}}(0,1)}}]
\end{equation}
In addition, step (c) will hold when ${\rm{E}}[{e^{\theta {Z_1}}}]{\rm{E}}{[^{ - \theta {Y_1}}}] \le {\rm{1}}$.

Hence, ${V_1},{V_2},{V_3}, \cdots ,{V_j}, \cdots ,{V_{k - 2}}$ form a non-negative supermartingale.
We further have
\begin{equation}
  \begin{aligned}
Pr\{B > y\}& \le \Pr \{ {\mathop {\max }\limits_{1 \le j \le k - 2} ( {\sum\limits_{n = j}^{k - 2} {[ {{Z_n} - {Y_{n + 1}}} ]} } ) > y} \}\\
 &= \Pr \{ {\mathop {\max }\limits_{1 \le j \le k - 2} {V_{j - i - 3}} > {e^{\theta y}}} \}\\
&\mathop  = \limits^a \Pr \{ {V_{j^*} > {e^{\theta y}}} \}\\
&\mathop  \le \limits^b {e^{ - \theta y}}{\rm{E[}}{V_{j^*}}{\rm{]}}\\
&\mathop  \le \limits^c {e^{ - \theta y}}{\rm{E[}}{V_1}{\rm{]}}\\
&= {e^{ - \theta y}}{\rm{E}}[{e^{\theta {Z_{k - j - 3}}}}]{\rm{E}}[{e^{ - \theta {Y_{k - j - 3}}}}]\\
&\mathop  \le \limits^d {e^{ - \theta y}}
  \end{aligned}
\end{equation}
Here, in step (a), $ V_{j^*} (1 \le j^* \le k-2) $ represents the maximal valve among $\{ {{V_1},{V_2},{V_3}, \cdots ,{V_j}, \cdots ,{V_{k - 2}}} \}$.
Using the Chernoff's inequality, we can complete the derivation of step (b).
In step (c), as ${V_1},{V_2},{V_3}, \cdots ,{V_j}, \cdots ,{V_{k - 2}}$ form a non-negative supermartingale, and based on Doob's inequality for submartingales and the formulation for supermartingales, there holds ${\rm{E[}}{V^*}{\rm{]}} \le {\rm{E[}}{V_1}{\rm{]}}$ \cite{doob1953stochastic}.
In addition, the definition of $V_1$ and the independence between $Y_k$ and $Z_k$ are contributed to step (d).

Hence, there holds
\begin{equation}
  \begin{aligned}
\Pr \{ B \le y\} \ge 1 - {e^{ - \theta y}}
  \end{aligned},
\end{equation}
which completes the proof. $\hfill\blacksquare$

Applying Lemma 2 to Eq. (12), the peak AoI violation probability finally holds as
\begin{equation}
\begin{aligned}
&\Pr \{ {P^{{\rm{AoI}}}}(k) > d\} \\
 &\le 1 - \int_0^d {\int_0^x {\Pr \{ B \le y\} {f_Y}(x - y)dy} {f_Z}(d - x)dx} \\
  &\le 1 - \int_0^d {\int_0^x {(1 - {e^{ - \theta y}}){f_Y}(x - y)dy} {f_Z}(d - x)dx} \\
 &= 1 - \int_0^d {(1 - {e^{ - \theta y}})[{f_Y}*{f_Z}(d - y)]dy}
\end{aligned}
\end{equation}
where $*$ denotes the convolution operator.

Note that in Eq. (21), $\theta$ is a non-negative parameter meeting ${\rm{E}}[{e^{\theta {Z_1}}}]{\rm{E}}{[^{ - \theta {Y_1}}}] \le {\rm{1}}$.
In addition, it is easily verified that $\Pr \{ {P^{{\rm{AoI}}}}(k) > d\}$ is a decreasing function in $\theta$.
Hence, the probabilistic bound of peak AoI can be tightened when $\theta$ is chosen according to the following expression
\begin{equation}
\begin{aligned}
\theta^{*}=\max\{\theta: {\rm{E}}[{e^{\theta {Z_1}}}]{\rm{E}}{[^{ - \theta {Y_1}}}] \le {\rm{1}}\}
\end{aligned}
\end{equation}

Also note that as the optimal $\theta$ is also related to the  characteristics of the inter-arrival time and that of service time, according to Eq. (22), the result of peak AoI violation probability in Eq. (21) can be applied to any scenario as long as the characteristics of the inter-arrival time and that of service time are available.
It is highlighted that those characteristics can be obtained under a given information update policy.

\section{Case Study}
In this section, the application of the derived result in Eq. (21) is illustrated with the help of two classical queueing models, i.e., the M/M/1 queueing model and the D/M/1 queueing model.

\subsection{M/M/1 Queuing Model}
In M/M/1 queuing model, the inter-arrival time and service time of each packet are i.i.d.
Let $\lambda$ denote the average inter-arrival time between any two adjacent packets and $\mu$ denote the average service time of a packet\footnote{Note: Here $\lambda$ and $\mu$ are defined to the time while not the rate.}, the PDF of the those two parameters holds as
\begin{equation}
\begin{aligned}
{f_Y}{\rm{(}}y{\rm{) = }}\frac{1}{\lambda }{e^{ - \frac{1}{\lambda }y}},~~{f_Z}{\rm{(}}x{\rm{) = }}\frac{1}{\mu }{e^{ - \frac{1}{\mu }x}}
\end{aligned}
\end{equation}

According to Eq. (21), the peak AoI violation probability of M/M/1 queueing model holds as

\begin{equation}
  \begin{aligned}
&\Pr \{ {P^{{\rm{AoI}}}}(k) > d\} ~~~~~~~~~~~~~~~~~~~~~~~~~~~~~~~~~~~~~~~~~~~~\\
& \le 1 - \int_0^d {(1 - {e^{ - \theta y}})[{f_Y}*{f_Z}(d - y)]dy} \\
& = 1 - \int_0^d {(1 - {e^{ - \theta (d - y)}})[{f_Y}*{f_Z}(y)]dy} \\
& = 1 - \frac{1}{{\mu  - \lambda }}\int_0^d {(1 - {e^{ - \theta (d - y)}})( {{e^{ - \frac{1}{\mu }y}} - {e^{ - \frac{1}{\lambda }y}}} )dy} \\
 &= \frac{{{\mu ^2}\theta }}{{( {\lambda  - \mu } )( {1 - \mu \theta } )}}{e^{ - \frac{1}{\mu }d}} - \frac{{{\lambda ^2}\theta }}{{( {\lambda  - \mu } )( {1 - \lambda \theta } )}}{e^{ - \frac{1}{\lambda }d}}\\
&~~~~~~~~~~~~~~~~~~~~~~~~~~~~~~~~{\rm{ + }}\frac{1}{{( {1 - \mu \theta } )( {1 - \lambda \theta } )}}{e^{ - \theta d}}
  \end{aligned}
\end{equation}
Here, $\theta$ can be optimized according to Eq. (22), there holds
\begin{equation}
  \begin{aligned}
{\rm{E}}[{e^{\theta {Z_1}}}]{\rm{E}}{[e^{ - \theta {Y_1}}}]=\frac{1}{{1 - \mu \theta }} \cdot \frac{1}{{1 + \lambda \theta }} \le {\rm{1}} \Rightarrow \theta^{*}  = \frac{{\lambda  - \mu }}{{\lambda \mu }}
  \end{aligned}
\end{equation}

Substituting Eq. (25) into Eq. (24), we further have
\begin{equation}
  \begin{aligned}
\Pr \{ {P^{{\rm{AoI}}}}(k) > d\} \le \frac{\lambda }{{2\mu  - \lambda }}{e^{ - \frac{{( {\lambda  - \mu } )d}}{{\lambda \mu }}}} + {e^{ - \frac{d}{\mu }}} - \frac{\lambda }{{2\mu  - \lambda }}{e^{ - \frac{d}{\lambda }}}
   \end{aligned}
\end{equation}

As a result, an upper bound on the average peak AoI for the case of M/M/1 is obtained as
\begin{equation}
  \begin{aligned}
&{\rm E}[ {{P^{{\rm{AoI}}}}(k)} ] = \int_0^\infty  {\Pr \{ {P^{{\rm{AoI}}}}(k) > d\} dd} \\
 & \le \int_0^\infty  {\frac{\lambda }{{2\mu  - \lambda }}{e^{ - \frac{{( {\lambda  - \mu } )d}}{{\lambda \mu }}}} + {e^{ - \frac{d}{\mu }}} - \frac{\lambda }{{2\mu  - \lambda }}{e^{ - \frac{d}{\lambda }}}dd}  \\
 &= \frac{{{\lambda ^2}}}{{( {\lambda  - \mu } )}} + \mu \;\;
  \end{aligned}
\end{equation}

\subsection{D/M/1 Queuing Model}
In D/M/1 queueing model, the inter-arrival time between two adjacent packets are deterministic while the service time of each packet follows the identically exponential distribution.
Let $D$ denote the inter-arrival time and $\mu$ denote the average service time of a packet. When $d>D$, the peak AoI violation probability of D/M/1 queueing model holds as
\begin{equation}
  \begin{aligned}
&\Pr \{ {P^{{\rm{AoI}}}} > d\}  \le 1 - \int_0^d {\Pr \{ B + {Y_k} \le x\} {f_Z}(d - x)dx} \\
&~~~~~~~~~~~~ = 1 - (0 + \int_D^d {(1 - {e^{ - \theta (x - D)}}) \cdot \frac{1}{\mu }{e^{ - \frac{1}{\mu }(d - x)}}dx} {\rm{)}}\\
&~~~~~~~~~~~~ = {e^{ - \frac{1}{\mu }(d-D)}} + \frac{1}{{1 - \theta \mu }}({e^{\theta (D - d)}} - {e^{ - \frac{1}{\mu }(d-D)}})
  \end{aligned}
\end{equation}
Here, $\theta$ can be optimized according to Eq. (22), there holds
\begin{equation}
  \begin{aligned}
{\rm{E}}[{e^{\theta {Z_1}}}]{\rm{E}}{[e^{ - \theta {Y_1}}}]=\frac{1}{{1 - \mu \theta }}{e^{ - \theta D}} \le {\rm{1}}
  \end{aligned}
\end{equation}
In this case, a closed-form optimal $\theta^{*}$ is unavailable while the analytical value can be obtained with the help of calculation tool like MATLAB.
Hence, while $d>D$, we finally have:
\begin{equation}
  \begin{aligned}
\Pr \{ {P^{{\rm{AoI}}}} > d\}  \le {e^{ - \frac{1}{\mu }(d - D)}} + \frac{{{e^{{\theta ^*}(D - d)}} - {e^{ - \frac{1}{\mu }(d - D)}}}}{{1 - {\theta ^*}\mu }}
  \end{aligned}
\end{equation}
where $\theta^*$ is the maximum allowable value satisfying Eq. (29).

As a result, an upper bound of the average peak AoI of D/M/1 queueing model can be obtained as
\begin{equation}
  \begin{aligned}
&{\rm{E[}}{P^{{\rm{AoI}}}}{\rm{]}} = \int_{\rm{0}}^D {\Pr \{ {P^{{\rm{AoI}}}} > d\} } dd + \int_D^\infty  {\Pr \{ {P^{{\rm{AoI}}}} > d\} } dd\\
&~~~~~~~~~= D + \int_D^\infty  {{e^{ - \frac{1}{\mu }(d - D)}} + \frac{{{e^{{\theta ^*}(D - d)}} - {e^{ - \frac{1}{\mu }(d - D)}}}}{{1 - {\theta ^*}\mu }}} dd\\
&~~~~~~~~~= D + \mu  + \frac{1}{\theta^* }
  \end{aligned}
\end{equation}

\section{Numerical Results}
In this section, numerical results are provided and discussed for the two cases, M/M/1 and D/M/1. The former M/M/1 case has often been  adopted in the literature for AoI performance analysis, e.g., see \cite{9324753}. The latter case D/M/1 corresponds to a practical setting where the packets are sent periodically.
The system utilization, defined as $\rho=\frac{\mathrm{E}[T^{\rm{S}}(k)]}{\mathrm{E}[T^{\rm{I}}(k-1,k)]}$, will also be used as a performance parameter.

\begin{figure}[t]
\centerline{\includegraphics[scale=0.55]{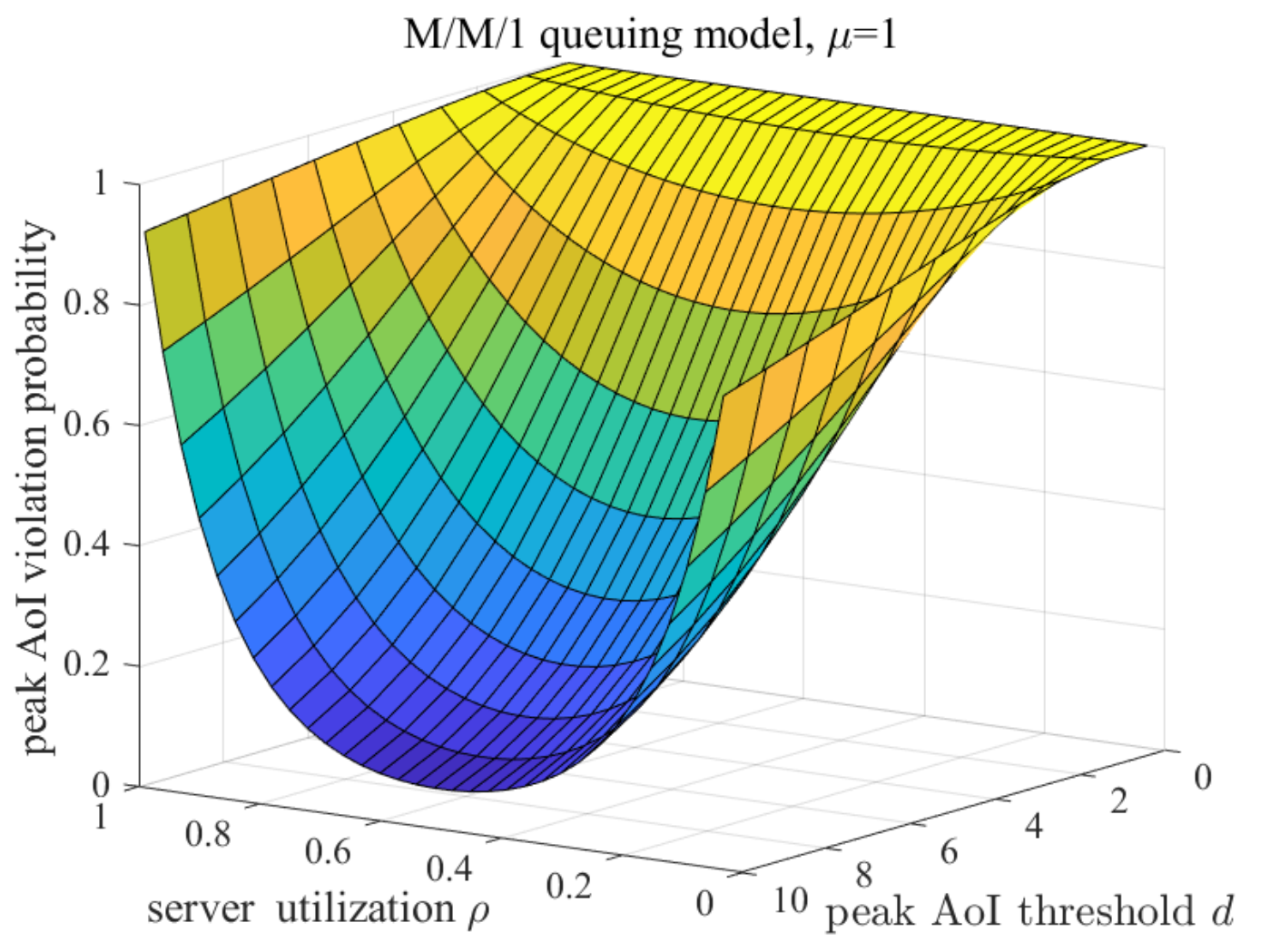}}
\caption{Peak AoI violation probability of the M/M/1 case.}
\end{figure}
\begin{figure}[t]
\centerline{\includegraphics[scale=0.55]{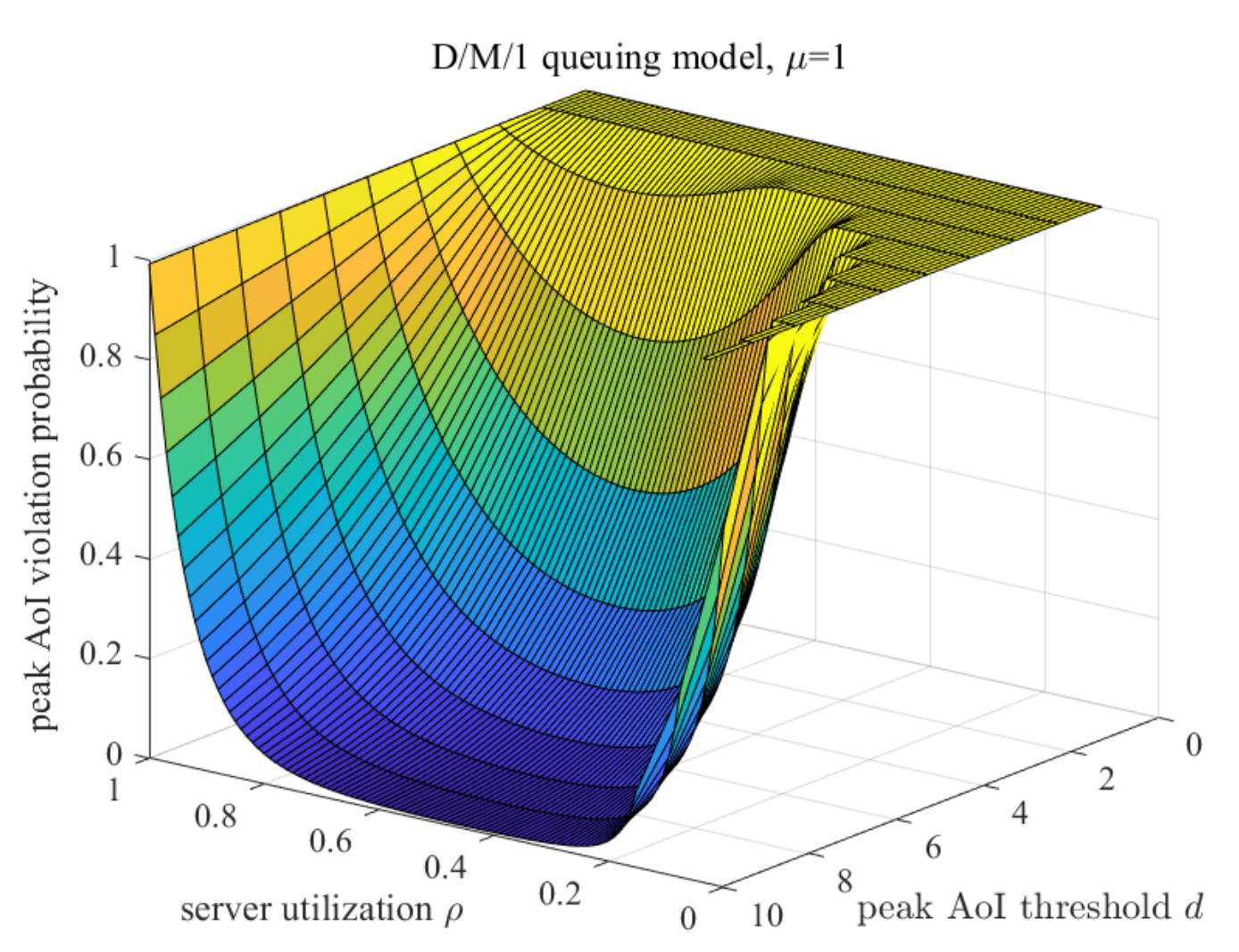}}
\caption{Peak AoI violation probability of the D/M/1 case.}
\end{figure}
Fig. 2 and Fig. 3 depict the peak AoI violation probability varying with the server utilization $\rho$ and the maximum tolerant threshold $d$ based on M/M/1 and D/M/1 queuing models, respectively.
We set the average service time for any packet as $\mu=1$ (time unit).
It is observed that the peak AoI violation probability decreases as the maximum tolerant threshold increase with a given server utilization.
Meanwhile, for any maximum tolerant threshold, there also exits an optimal server utilization setting to minimize the peak AoI violation probability.
This is because the peak AoI depends on inter-arrival time, queueing time and service time.
With a given service capability, the statistical characteristics of the service time is usually fixed.
In this regard, low server utilization means huge inter-arrival time.
Too much high inter-arrival time may result in few packet updated to the receiver, which deteriorates the peak AoI performance.
On the other hand, too much small inter-arrival time may result in high queuing delay for each packet, which also has negative impact on the peak AoI performance.
Therefore, it is critical to design an appropriate server utilization scheme for peak AoI guarantee.

\begin{figure}[t]
\centerline{\includegraphics[scale=0.6]{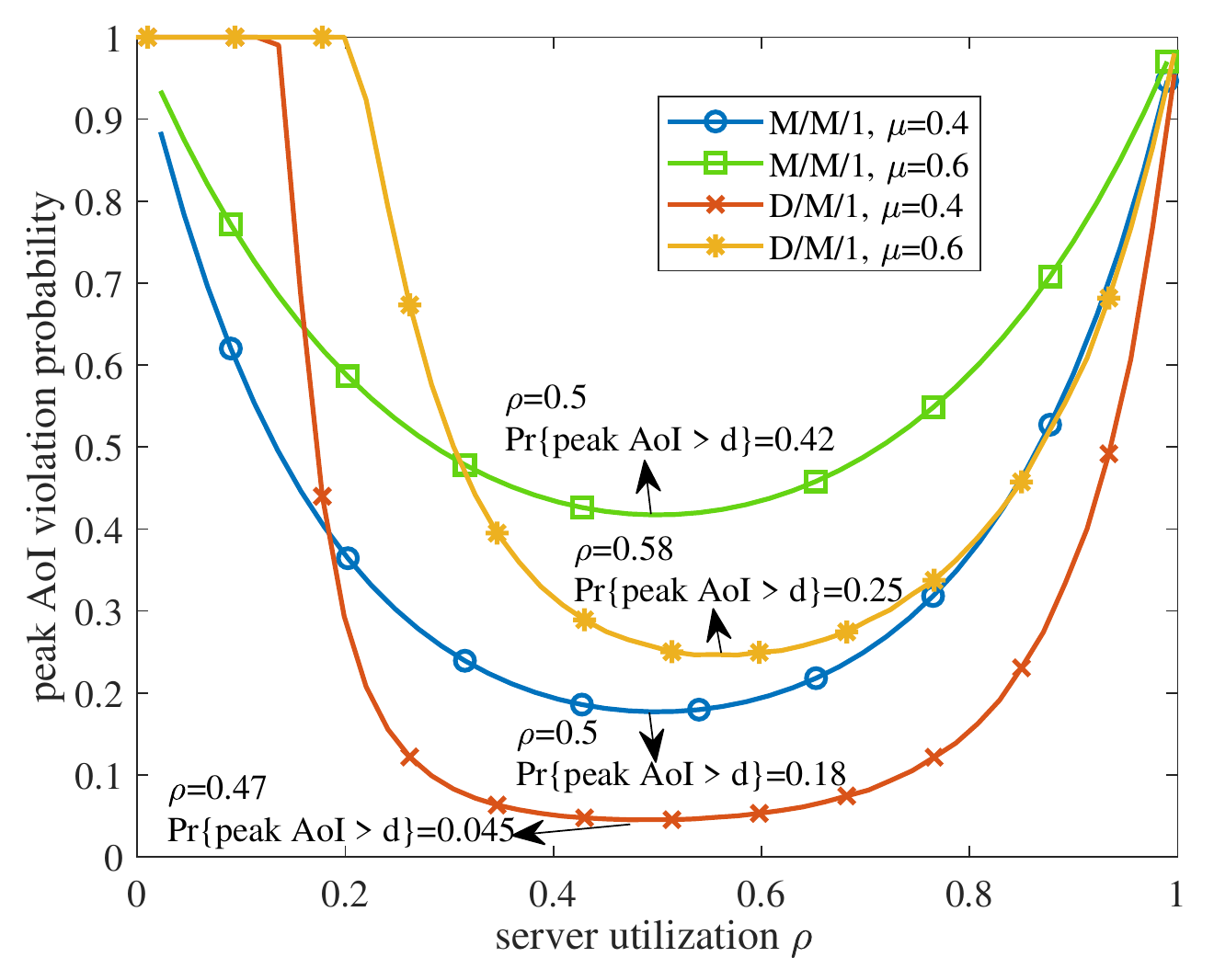}}
\caption{The influence of server utilization on the peak AoI violation probability.}
\end{figure}
Fig. 4 depicts the relationship between peak AoI violation probability and server utilization under different service capability.
We set the threshold of peak AoI as $d=3$ (time unit).
In addition to the observation from Fig. 2 and Fig. 3, it is found that the optimal server utilization of M/M/1 is irrelevant with the service capability.
Specifically, the optimal utilization is also equal to 0.5.
Differently, the optimal server utilization of D/M/1 queuing model is positively correlated with the average service time $\mu$.
Additionally, Fig. 4 verifies that the peak AoI violation probability can be reduced by decreasing the service time of a packet, i.e., improve the service capability.

\begin{figure}[t]
\centerline{\includegraphics[scale=0.6]{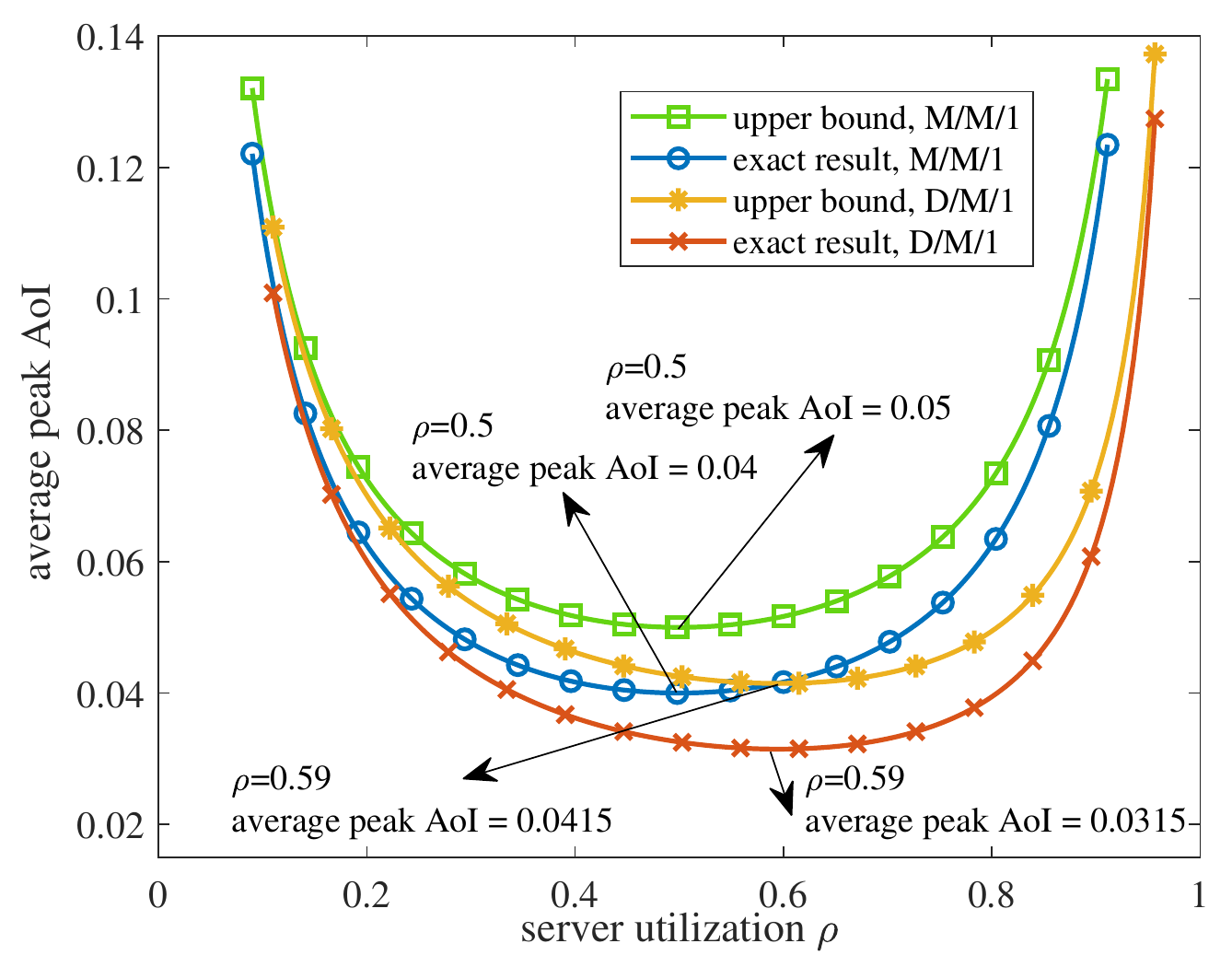}}
\caption{The influence of server utilization on the average peak AoI.}
\end{figure}
Fig. 5 depicts the relationship between the average peak AoI and the server utilization.
In addition, the obtained performance bound is compared with results from classical queueing theory analysis.
We set the service time of the system as $\mu=0.01$.
It is observed that there exists optimal configuration of server utilization in average peak AoI minimization for each case.
Additionally, it can be found that the obtained performance bounds for the M/M/1 case and D/M/1 case are closed to the corresponding exact result, respectively.
Specifically, the gaps between the upper bound and exact result for both cases are always a service time $\mu$.

\section{Conclusions}
In this paper, the probabilistic characteristics of the peak AoI have been studied.
Different from the literature where the focus is mostly on average (peak) AoI, we provided an analysis on peak AoI violation probability.
By decoupling the inter-arrival time and the sojourn time of each packet, the peak AoI can be expressed with the inter-arrival time and the service time of each packet.
With the help of martingale theory, an upper bound on peak AoI violation probability was derived for the general GI/GI/1 setting, and applied to two specific cases, namely  M/M/1 and D/M/1.
The impact of server utilization on peak AoI performance was also  investigated. Additionally, numerical results verifying the validity of the proposed peak AoI bound were presented and discussed.

\section{Acknowledgement}
This work was supported in part by the National Natural Science Foundation of China under grants 61901078, 61871062 and U20A20157, and in part by Natural Science Foundation of Chongqing under grant cstc2020jcyj-zdxmX0024, and in part by  University Innovation Research Group of Chongqing under grant CXQT20017, and in part by the China University Industry-University-Research Collaborative Innovation Fund (Future Network Innovation Research and Application Project) under grant 2021FNA04008,

\bibliographystyle{IEEEtran}
\bibliography{bibfile}

\end{document}